\documentclass[11pt]{article}
\usepackage{geometry}                
\usepackage{graphicx}
\usepackage{amssymb}
\usepackage{caption}
\usepackage{subcaption}
\usepackage{amsmath}
\usepackage{siunitx}
\usepackage{fixmath}
\usepackage{MnSymbol}
\usepackage{amsfonts}
\usepackage{array}
\usepackage{multirow}
\DeclareMathAlphabet{\mathpzc}{OT1}{pzc}{m}{it}

\usepackage{caption}
\usepackage{subcaption}

\title{Vertically Driven Waves: Energy Transfer Between Gravity Waves Revisited}
\author{Clifford Chafin\\\ \small{Department of Physics, North Carolina State University, Raleigh, NC 27695} \thanks{cechafin@ncsu.edu}}

\begin{document}
\maketitle
\begin{abstract}
We investigate the energy transfer from large waves to small ones through vertical acceleration and demonstrate that this is a much larger effect than that of the potential energy changes of the small waves moving over the larger ones.  Rates of exponential growth for this process are given and limits on the stable size of small waves in the horizontal accelerations from the larger ones are derived.  We discuss the possibility of this being a manifestation of the Benjamin-Feir instability.  
\end{abstract}

\section{Introduction}
Small waves moving over large ones exchange energy.  Hasselmann predicted that mass and momentum would flow to the smaller ones \cite{Ha71}.  This analysis uses mass flux and radiation stress analysis based on earlier work \cite{LH69, LH64}.  It is known that waves tend to have Eulerian fluxes that cancel net drift and the role of surface flows can significantly alter wave dynamics \cite{Sm06} so one might wonder if this would alter this analysis.  Recent work by this author has reconsidered the role of radiation stress as a pseudostress that need not result in real forces and that there is no substitute for the true long range nonlinear pressure forces in the analysis \cite{Chafin-rogue, Chafin-acoustic}.  

In the derivation of wave interactions one often starts with a linearity assumption and introduces ``interactions'' via the nonlinear advective term.  When we define ``small'' for periodic waves we use the criterion $a/\lambda<<1$.  In the N-S equations, when the nonlinear term $\nabla\Phi\cdot\nabla\Phi$ is smaller than $\partial_{t}\Phi$ and $P$ everywhere in the solution this is satisfied.  However, we cannot assume that the linear superposition of two such ``small'' waves keeps this nonlinear term negligible.  When we superimpose two small periodic waves given by $(A,\Lambda)$ and $(a,\lambda)$, with $A\gg a$ and $\Lambda\gg\lambda$ but $A\sim\lambda$ we get a cross term that is does not let us decompose the terms into two independent equations.  This puts limits of when two ``small'' waves may be reasonably superimposed for the purpose of interactions.

However, there are even further complications.  Small waves tend to ride over the tops of large ones treating them as the new equilibrium surface.  However if we add the amplitudes $A+a$ then we cannot add the velocity potentials as $\Phi+\phi$ because the exponential increase in $\phi$ over the added height region of the higher wave gives a much too large kinetic energy.  

If we use the picture of packet of small waves using the large ones as the equilibrium surface then we have the problem that the drift of a right moving packet gives a clockwise angular momentum in the trough and a counterclockwise contribution at the crest of the larger wave.  The angular momentum of larger waves is disproportionately greater so shape changes in it can compensate for this but it is still a point of frustration to resolve this by a simple perturbative approach.  

This state of affairs seems plenty to justify an independent treatment of this problem.  We will completely neglect the effect of small waves ramping up and down larger waves and instead focus on the effect of vertical acceleration.  This is most appropriate since slow rise up the slope compared to vertical acceleration $\frac{1}{2}\sqrt{g/k}A/\Lambda\ll A\Omega$ follows from $\frac{\lambda}{\Lambda}\ll1$.  We will show that energy is imparted to small waves using simple conservation law arguments and give a simple expression for the rate.  

%
%

\section{Vertically Driven Waves}\label{Vertical}
We generally consider waves in a medium with constant restoring force.  This is equivalent to a system with uniform acceleration.  The classical jerk, $\dot{\text{a}}$, of the system tells us how much this is changing.  The Faraday instability is the formation of wave patterns as a surface is vertically oscillated.  Here we investigate extant surface waves under a $\delta$-function jerk.  This may seem a little artificial but we are interested in how a large wave's vertical motion affects small waves on top of them.  There are other cases where the restoring forces are driven by surface tension, as in capillary waves, or other properties of the medium that can be more readily changed.  Let us investigate the following case with an exact solution.  

\begin{figure}[!ht]
   \centering
   \includegraphics[width=2in,trim=0mm 90mm 0mm 30mm,clip]{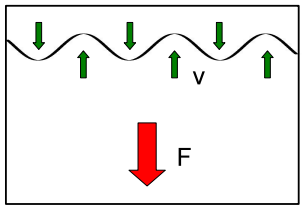} 
   \caption{A vertically accelerated container with a standing wave.}
   \label{standing}
\end{figure}

Consider a standing wave in a container in which we can suddenly change the force of gravity, $g_{eq}\rightarrow g_{1}$, as in Fig.~\ref{standing}.  By sudden we mean that the change in the force happens much faster than the period of the waves $\partial_{t} \text{ln}(g)>>\omega$.  This could be done by accelerating the container vertically.\footnote{We could envision a version of this for capillary waves by applying a surfactant.  An analogous effect, driven by different surfactant physics, is discussed in \cite{Levy06}.  Since we still are using the small wave approximation we must maintain that $a\omega^{2}<<g_{1}$ for the resulting waves.  }  Over one cycle, standing waves transfer energy from purely kinetic to purely potential energy and back again.  This means that the exact time we change the force affects the energy change in the wave.  Since the linear and angular momentum of a standing wave is zero, we can compute the evolution based on energy conservation.  If we turn it on when the surface is flat, then energy conservation, $\frac{1}{2}\rho g_{eq}a_{0}^{2}=\frac{1}{2}\rho g_{1}a_{1}^{2}$, gives the new wave amplitude $a_{1}=\sqrt{\frac{g_{eq}}{g_{1}}}a_{0}$ and the new period is $\omega_{1}=\sqrt{g_{1}k}$.  If we turn it on at the wave maximum,  the energy density increases $\mathcal{E}=\frac{1}{2}\rho g_{eq}a_{0}^{2}\rightarrow\frac{1}{2}\rho g_{1}a_{0}^{2}$ and the maximum amplitude is unchanged.

\begin{figure}[!ht]
   \centering
   \includegraphics[width=2in,trim=0mm 0mm 0mm 130mm,clip]{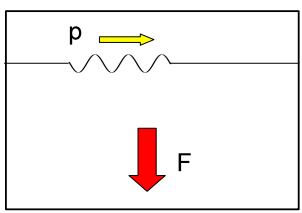} 
   \caption{The same container with a traveling wave. }
   \label{traveling}
\end{figure}

Now let us consider the case of a traveling wave as in Fig.~\ref{traveling}.  The linear and angular momentum densities are nonzero and the KE and PE of the wave remain exactly half the total energy at all times.  This means the time that we turn on the extra force is no longer relevant and the PE of the wave is always altered by the new force.  

We evaluate the initial energy and momenta densities for the wave 
\begin{align}
\mathcal{E}=\mathcal{K}+\mathcal{U}&=\frac{1}{4}\rho g_{eq}a_{0}^{2}+\frac{1}{4}\rho g_{eq}a_{0}^{2}\\
\mathpzc{p}&=\frac{1}{2}\rho a_{0}^{2}\sqrt{{g_{eq}}{k}}\\
\mathcal{L}&=-\frac{1}{4}\rho a_{0}^{2}\sqrt{\frac{g_{eq}}{k}}
\end{align}
The energy is altered by the new force to 
\begin{align}\label{shiftedE}  
\mathcal{E}\rightarrow\mathcal{E}_{i}=\frac{1}{4}\rho g_{eq}a_{0}^{2}+\frac{1}{4}\rho g_{1}a_{0}^{2}
\end{align}
leaving $\mathpzc{p}$ and $\mathcal{L}$ unchanged.  
  
Linear theory lets us specify the new $\Phi$ and $\eta$ at the start time and evolve using the new $\omega$ corresponding to the new $g_{1}$ and old $k$.  This theory predicts that we will have only forwards and backwards waves with wavevector $k$ in the solution.  Applying this fact to our conservation laws we can derive the new amplitudes of the resulting waves.

To get conservation of energy momentum with the new wave we need a backwards traveling wave.  The resulting energy and momenta of these two product waves are
\begin{align}
\mathcal{E}_{f}&=\frac{1}{2}\rho g_{1}a_{1}^{2}+\frac{1}{2}\rho g_{1}a_{2}^{2}\\
\mathpzc{p}_{f}&=\frac{1}{2}\rho a_{1}^{2}\sqrt{{g_{1}}{k}}-\frac{1}{2}\rho a_{2}^{2}\sqrt{{g_{1}}{k}}\\
\mathcal{L}_{f}&=-\frac{1}{4}\rho a_{1}^{2}\sqrt{\frac{g_{1}}{k}}+\frac{1}{4}\rho a_{2}^{2}\sqrt{\frac{g_{1}}{k}}
\end{align}
Only the energy and linear momentum equations are independent here.  Setting the initial and final energies and momenta equal we find 
\begin{align}
a_{1}&=\left(\frac{a_{0}}{2}\right)\left( 1+\sqrt{\frac{g_{0}}{g_{1}}} \right) \\
a_{2}&=\left(\frac{a_{0}}{2}\right)\left( 1-\sqrt{\frac{g_{0}}{g_{1}}} \right) 
\end{align} 
where we have set $g_{0}=g_{eq}$ for the moment.  

%


Instead of our abrupt change which creates oppositely moving packets, we now can consider the packet to bifurcate but still be overlapping (except for some small end contributions).  We could try to choose the optimal moments to accelerate to gain the most energy and decelerate to loose the least.  This would give an upper bound.  Instead let us assume a random phase result and iterate eqn.~\ref{shiftedE} to get an idea of the typical effects on a wave.  Assuming that we alternate equally between $g_{0}$ and $g_{1}$ (so $g_{1}=g_{eq}+\delta$ and $g_{0}=g_{eq}-\delta$) we find that the amplitude of the larger (dominant) contribution after $n$ iterations is
\begin{align}
a_{n}=\frac{1}{4}\left(2+{\frac{g_{0}+g_{1}}{\sqrt{g_{0}g_{1}}}}\right)^{n}a_{0}
\end{align}
where the argument is noted to be the ratio of the arithmetic and geometric means.  
Therefore, $a_{n}=a_{0}(1+\frac{n}{4g_{eq}^{2}}\delta^{2}-\ldots)$ which shows that we are pumping energy into the wave.  Since $k$ is unchanged by this action and $ak$ is growing this eventually leads to wave instability so that breaking waves impart momentum to the surface flow.  Assuming the vertical acceleration is from a large wave, $(A, \Lambda)$, so that a~$=A\Omega^{2}$ and the rate of iteration is $\dot{n}=\frac{\Omega}{2\pi}$ and we have
\begin{align}
\dot{a}=\frac{\Omega}{8\pi} (AK)^{2}a=\tau^{-1} a
\end{align}
so that the small waves grow exponentially with time.  The maximum slope for a wave before it breaks is when $ak\approx 0.44$.  We have
\begin{align}
(ak)=(ak)_{0} e^{t/\tau}
\end{align}
Our approximations are not valid for nonlinear waves but they can give us an order of magnitude for the time it takes for a wave to increase its slope to the point of breaking as $(\ln(0.44)-\ln(a_{0}k))\tau \approx t_{b}$.  It is interesting that this growth is exponential as in the case of the Benjamin-Feir instability \cite{bf67,benjamin67} but only seems to lead to pumping energy in the shorter wave direction and the group velocity is not anomalous.  However, it is possible that there is a nonlinearity hidden in this analysis that is not obvious and makes this equivalent to B-F.  This is not uncommon for results based on conservation laws where, for example, we can easily compute the momentum of Stokes drift for an Airy wave even though the drift itself is typically considered a nonlinear correction.

%
%
%
%
%

\section{Horizontal Acceleration}


Small waves that travel over large ones experience both a vertical acceleration and a change in the local velocity of the underlying flow.  This will stretch and contract the wave train.  Additionally, waves will travel into regions of different underlying flow.  They will undergo a change in forwards of velocity of $\Delta v=2A\Omega$.  

%

\begin{figure}[!ht]
   \centering
   \includegraphics[width=2in,trim=0mm 30mm 0mm 150mm,clip]{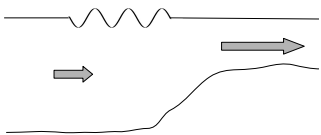} 
   \caption{A small wave moving over an accelerating flow.}
   \label{wavesonshear}
\end{figure}

To investigate this let us consider a simpler example, a flow that is being vertically compressed to alter the underlying velocity field that the waves ride across as in Fig.~\ref{wavesonshear}.  Let us consider this simple case where the acceleration in localized so that we begin and end at a constant mean flow with a change in velocity $u_{f}-u_{i}=u$.
Conservation of mass flux tell us that when the flow is backwards, $u<0$, and the group velocity is less than the change in flow $v_{g}<|u|$ then the waves must undergo some critical steepening and break.  This implies no small waves with $v_{g}<A\Omega$ can exist.  
Combining this with the previous results we see that the small waves must have wavelength $\lambda>8\pi (AK)A$ and that these waves will grow towards breaking with a time constant $\tau^{-1}\sim(AK)^{2}\sqrt{K}$.

\section{Conclusions}
We have investigated the role of large waves when acting on a set of much smaller ones from the point of view of the vertical and horizontal accelerations they undergo rather than the elevation changes the small waves undergo traversing the larger ones.  This is shown to give the dominant effect.  These give interesting bounds on the minimum wavelength of small waves to survive one period of the larger waves' motion and a rate that small waves pull energy from the large ones.  Interestingly, this grows exponentially but is not clearly connected with the Benjamin-Feir instability and seems to only send energy to shorter wavelengths.  This seems worth considering further to see if it is just an approximate form of B-F or a new effect entirely.  

The correct way to do this problem analytically is unclear for reasons stated in the introduction.  Radiative stress is a tempting tool since it involves depth averaging and hides the problems alluded with superposition however this seems like a superficial fix and these sorts of stresses don't necessarily transfer energy in this fashion.  A future direction might be to keep track of the surface forces the small waves directly exert on the surface of the big ones and vice versa so that all the conserved quantities are manifest.  This may involve some active deformation in wave shape that must be included.  It does have the distinct advantage that we can tell how long waves can be reduced in size by surface forces as in wave sheltering.  The small waves have plenty of linear momentum per unit of energy to absorb the momentum of the damping large waves but where the angular momentum (measured relative to a point on the equilibrium surface) goes is problematic.  The only option is for end-of-packet losses which means an infinite wave train analysis will not be sufficient.  

\section*{Acknowledgements}
Thanks to Renee Foster and Bo Hemphill for proofing this and earlier versions.  Special thanks to Thomas Sch\"{a}fer for thoughtful discussions.

\end{document}